\documentclass[aps,prl,preprint,superscriptaddress]{revtex4-2}
\usepackage{tabularray}
\usepackage{amsmath}
\usepackage{makecell}
\usepackage{graphicx}
\usepackage{dcolumn}
\usepackage{bm}

\usepackage[utf8]{inputenc}
\usepackage{babel}
\usepackage[T1]{fontenc}
\usepackage{lmodern,microtype,ellipsis}

\usepackage{siunitx,tcolorbox,lastpage,amsmath,csquotes}
\usepackage[colorlinks=true,linkcolor=black,citecolor=black,urlcolor=black]{hyperref}
\usepackage{physics,scalerel,amssymb}

\usepackage{color}
\begin{document}
	
	\title{Influence of excitation energy on microscopic quantum pathways for ultrafast charge transfer in van der Waals heterostructures}
	
	
	
	\author{Niklas Hofmann}
	\affiliation{Department for Experimental and Applied Physics, University of Regensburg, 93053 Regensburg, Germany}
	
	\author{Johannes Gradl}
	\affiliation{Department for Experimental and Applied Physics, University of Regensburg, 93053 Regensburg, Germany}
	
	\author{Leonard Weigl}
	\affiliation{Department for Experimental and Applied Physics, University of Regensburg, 93053 Regensburg, Germany}

	\author{Stiven Forti}
	\affiliation{Center for Nanotechnology Innovation@NEST, Istituto Italiano di Tecnologia, Pisa, Italy}
	\affiliation{Graphene Labs, Istituto Italiano di Tecnologia, Genova, Italy}
	
	\author{Camilla Coletti}
	\affiliation{Center for Nanotechnology Innovation@NEST, Istituto Italiano di Tecnologia, Pisa, Italy}
	\affiliation{Graphene Labs, Istituto Italiano di Tecnologia, Genova, Italy}
	
	\author{Isabella Gierz}
	\email[]{isabella.gierz@ur.de}
	\affiliation{Department for Experimental and Applied Physics, University of Regensburg, 93053 Regensburg, Germany}
	

	\date{\today}

	\begin{abstract}

	Efficient charge separation in van der Waals (vdW) heterostructures is crucial for optimizing light harvesting and detection applications. However, precise control over the microscopic pathways governing ultrafast charge transfer remains an open challenge. These pathways are intrinsically linked to charge transfer states with strongly delocalized wave functions that appear at various momenta in the Brillouin zone. Here, we use time- and angle-resolved photoemission spectroscopy (trARPES) to investigate the possibility of steering carriers through specific charge transfer states in a prototypical WS\textsubscript{2}-graphene heterostructure. By selectively exciting electron-hole pairs at the K-point (A-exciton resonance) and close to the Q-point (C-exciton resonance) of WS\textsubscript{2} with different pump photon energies, we find that charge separation is faster at higher excitation energies. This behavior is attributed to the fact that absorption at the C-exciton resonance generates electron-hole populations at energies well above the direct band gap. The resulting elevated carrier temperatures open an additional, highly efficient charge-transfer channel for holes in the WS\textsubscript{2} valence band, leading to an overall acceleration of interlayer hole transfer for C-exciton excitation. The microscopic insights gained in this work can be leveraged to optimize the performance of vdW heterostructures in optoelectronic devices.

	\end{abstract}


	\maketitle

\section{Introduction}

The vast selection of 2D materials available today, such as graphene and monolayer transition metal dichalcogenides, enables the design of novel heterostructures with tailored properties \cite{Geim2013, Novoselov2016, Jin2018}. Such heterostructures commonly feature efficient absorption of visible light followed by ultrafast charge separation \cite{He2014a, Massicotte2016, He2017, Ji2017, Song2018, Yuan2018, Fu2021} with great potential for applications in light harvesting and detection. The driving force behind ultrafast charge separation is the band alignment, where photoexcited electrons and holes rapidly relax to the conduction band minimum and valence band maximum in separate layers. Charge transfer states, formed by hybridization between the vdW-coupled layers, create delocalized wave functions that serve as ultrafast tunneling channels \cite{Zheng2017, Liu2022, Hofmann2023}. These states occur at various momenta and energies in the Brillouin zone but do not contribute equally to charge separation. Based on existing models \cite{Zheng2017, Hofmann2023, Wang2016, Long2016, Li2017, Liu2021, Krause2020}, the efficiency of different tunneling channels depends on the strength of the hybridization, the size of the energy barriers that photo-excited carriers must overcome to reach the charge transfer state, as well as the available tunneling phase space. The possibility to steer carriers through specific charge transfer states, e.g. by selectively generating them at particular momenta and with controlled excess energies, remains unexplored.

Here, we investigate this concept in a prototypical vdW heterostructure consisting of monolayer WS\textsubscript{2} and graphene. We photoexcite the heterostructure with pump photon energies of $\hbar\omega=2.0\,\mathrm{eV}$ and $\hbar\omega=3.1\,\mathrm{eV}$ to excite electron-hole pairs at the K-point (A-exciton) and between the $\Gamma$- and Q-point (C-exciton) of WS\textsubscript{2}, respectively, and probe the charge transfer dynamics directly in the band structure using time- and angle-resolved photoemission spectroscopy (trARPES) \cite{Hofmann2025}. We find that charge separation is significantly faster following $\hbar\omega=3.1\,\mathrm{eV}$ excitation, which we attribute to the activation of a second, highly efficient charge transfer channel in the valence band. Our findings introduce new possibilities for optimizing charge separation in vdW heterostructures, paving the way for more efficient light-harvesting and detection technologies.

\section{Methods}

\textbf{Sample growth}: Pretreated 4H-SiC substrates were H-etched to remove scratches and graphitized in Ar atmosphere. The resulting carbon monolayer with $\left(6\sqrt{3}\times6\sqrt{3}\right)\mathrm{R}30^{\circ}$ structure was decoupled from the SiC substrate by H-intercalation, yielding quasi-freestanding monolayer graphene \cite{Riedl2009}. WS\textsubscript{2} was then grown by chemical vapour deposition (CVD) from solid WO\textsubscript{3} and S precursors \cite{Rossi2016, Forti2017}. Both graphene and WS\textsubscript{2} growth were monitored using Raman spectroscopy, atomic force microscopy and secondary electron microscopy. WS\textsubscript{2} was found to grow in the shape of triangular islands with side lengths in the range of $300-700\,\mathrm{nm}$ with twist angles of either 0° or 30° with respect to the graphene layer \cite{Hofmann2023}.

\textbf{trARPES}: The setup was based on a Ti:Sa amplifier (Astrella, Coherent) with $1\,\mathrm{kHz}$ repetition rate, $35\,\mathrm{fs}$ pulse duration, $1.55\,\mathrm{eV}$ photon energy, and $7\,\mathrm{mJ}$ output power. Of this, $2\,\mathrm{mJ}$ were frequency-doubled and used to drive high harmonics generation (HHG) in an Argon gas jet. The 7th harmonic at $21.7\,\mathrm{eV}$ was isolated using a grating monochromator and focused onto the sample with a beam diameter of $250\,\mathrm{\mu m}$. A small part of the $2\,\mathrm{mJ}$ output was frequency doubled to generate pump pulses with a photon energy of $3.1\,\mathrm{eV}$ with a fluence of $0.4\,\mathrm{mJ}/\mathrm{cm}^2$ at the focus. The remaining $5\,\mathrm{mJ}$ seeded an optical parametric amplifier (Topas Twins, Light Conversion). One of the signal beams was frequency doubled to generate pump pulses with a photon energy of $2.0\,\mathrm{eV}$ with a fluence of $1.7\,\mathrm{mJ}/\mathrm{cm}^2$ at the focus. The photoemitted electrons were detected with a hemispherical analyzer (Phoibos 100, SPECS) to obtain snapshots of the occupied band structure along the $\Gamma$K direction of graphene. The temporal and energy resolutions were $\sim200\,\mathrm{meV}$ and $\sim160\,\mathrm{fs}$, respectively.

\section{Results}

\begin{figure}
		\includegraphics[width = \columnwidth]{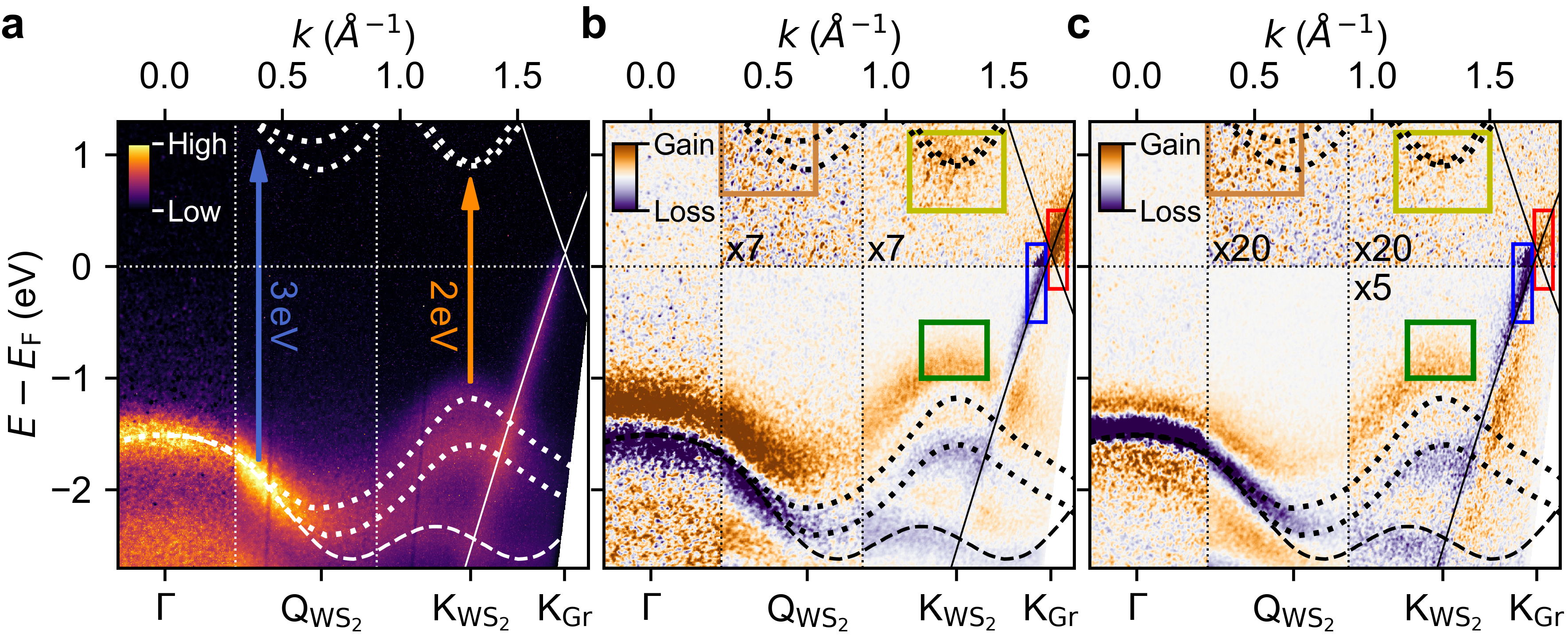}
		\caption{\textbf{trARPES data of WS\textsubscript{2}-graphene heterostructure.} \textbf{a)} ARPES snapshots for negative pump-probe delay taken along the $\Gamma$K direction of graphene. Orange and blue arrows indicate the two different excitation schemes for $\hbar\omega=2.0\,\mathrm{eV}$ and $\hbar\omega=3.1\,\mathrm{eV}$, respectively. Dotted, dashed and continuous lines represent the theoretical band structures for WS\textsubscript{2} with twist angles of 0° and 30° \cite{Zeng2013} and graphene \cite{Wallace1947}, respectively. \textbf{b)} Pump-induced changes of the photocurrent in a) 240fs after excitation with $\hbar\omega=2.0\,\mathrm{eV}$. \textbf{c)} Same as b) but $310\,\mathrm{fs}$ after excitation with $\hbar\omega=3.1\,\mathrm{eV}$. Colored boxes in b) and c) mark the regions of integration for the pump-probe traces displayed in Fig. \ref{figure2}.}
		\label{figure1}
	\end{figure}

Figure \ref{figure1}a shows an ARPES image of the WS\textsubscript{2}-graphene heterostructure at negative pump-probe delay before the arrival of the pump pulse. The dotted, dashed, and continuous lines indicate the theoretical band structures of WS\textsubscript{2} islands with a twist angle of 0°, WS\textsubscript{2} islands with a twist angle of 30° \cite{Zeng2013} and graphene \cite{Wallace1947}, respectively. The bands are shifted in energy to fit the experimentally observed band alignment and doping level. Orange and blue arrows mark the electronic transitions in WS\textsubscript{2} excited at $\hbar\omega=2.0\,\mathrm{eV}$ and $\hbar\omega=3.1\,\mathrm{eV}$, respectively \cite{Magnozzi2020}. Here, we will only consider WS\textsubscript{2} islands with 0° twist angle. The influence of the twist angle on ultrafast charge separation will be investigated in a separate work.

Figure \ref{figure1}b depicts the pump-induced changes of the photocurrent in Fig. \ref{figure1}a $240\,\mathrm{fs}$ after photoexcitation at $\hbar\omega=2.0\,\mathrm{eV}$ with a fluence of $F=1.7\,\mathrm{mJ}/\mathrm{cm}^2$. Brown and blue colors indicate a gain and loss of photoelectrons, respectively, with respect to negative pump-probe delay. Figure \ref{figure1}c is the same as Fig. \ref{figure1}b but for photoexcitation at $\hbar\omega=3.1\,\mathrm{eV}$ with a fluence of $F=0.4\,\mathrm{mJ}/\mathrm{cm}^2$ and $310\,\mathrm{fs}$ after photoexcitation. For both excitation energies we observe gain in the conduction band (CB) of WS\textsubscript{2}, gain (loss) above (below) the equilibrium Fermi level in graphene, and a complex gain-loss signal in the WS\textsubscript{2} valence band (VB) with contributions from band shifts and broadening as previously discussed in \cite{Krause2020, Hofmann2023}. 

To gain access to the transient carrier population of different parts of the band structure, we integrate the photocurrent over the areas marked by the colored boxes in Figs. \ref{figure1}b and c. The corresponding pump-probe traces are shown in Fig. \ref{figure2} together with single-exponential fits (see supplemental material) yielding the decay times $\tau$.  Dashed vertical lines in Fig. \ref{figure2} indicate the pump-probe delay $t_\mathrm{ex}$ where the respective pump-probe trace reaches its extremum. The fit results are summarized in Table \ref{table1}. The charge carrier dynamics inside the Dirac cone (Fig. \ref{figure2}a) are found to be quite similar for $\hbar\omega=2.0\,\mathrm{eV}$ and $\hbar\omega=3.1\,\mathrm{eV}$, respectively, considering that the pump-pulse cross-correlation (grey-shaded areas in Fig. \ref{figure2}) was longer for $\hbar\omega=3.1\,\mathrm{eV}$ ($90\,\mathrm{fs}$) than for $\hbar\omega=2.0\,\mathrm{eV}$ ($50\,\mathrm{fs}$). The population dynamics of the WS\textsubscript{2} bands, however, show important differences when comparing the two excitation energies. We find that the gain above the equilibrium position of the WS\textsubscript{2} VB (Fig. \ref{figure2}b) reaches its maximum $\sim250\,\mathrm{fs}$ later for $\hbar\omega=2.0\,\mathrm{eV}$ than for $\hbar\omega=3.1\,\mathrm{eV}$. Further, for $\hbar\omega=2.0\,\mathrm{eV}$, the pump-probe signal of the WS\textsubscript{2} CB at K is much bigger than close to Q, indicating that the photoexcited electrons are mainly confined to the K valley (Fig. \ref{figure2}c1). For $\hbar\omega=3.1\,\mathrm{eV}$, however, we find that the population of the WS\textsubscript{2} CB at K is smaller than the one close to Q and reaches its maximum $\sim130\,\mathrm{fs}$  later (Fig. \ref{figure2}c2).

To gain direct access to the timescales for ultrafast charge separation, we extract the related charging-induced band shifts of the WS\textsubscript{2} VB and CB as follows \cite{Krause2020, Hofmann2023}. First, we extract energy distribution curves (EDCs) at selected momenta from the trAPRES snapshots in Fig. \ref{figure1} that we fit with an appropriate number of Gaussian peaks (see supplemental material) to obtain the transient binding energies shown in Fig. \ref{figure3}a. Next, we subtract the transient position of the WS\textsubscript{2} VB from the transient position of the WS\textsubscript{2} CB yielding the transient band gap $E_\mathrm{gap}$ shown in Fig. \ref{figure3}b. Finally, assuming that the band gap changes symmetrically around its center, we add (subtract) half of the transient band gap change $\Delta E_\mathrm{gap}/2$ to (from) the transient position of the WS\textsubscript{2} VB (CB) to obtain the shifts $\Delta E_\mathrm{charge}^\mathrm{WS\textsubscript{2}}$ in Fig. \ref{figure3}c that we previously attributed to a transient charging of the WS\textsubscript{2} layer with excess electrons \cite{Krause2020, Hofmann2023}. 

The corresponding charging shift of the graphene layer in Fig. \ref{figure3}d is obtained by extracting momentum distribution curves (MDCs) that we fit with a Lorentzian (see supplemental material). The resulting peak positions are averaged and multiplied with the slope of the Dirac cone to yield the data points in Fig. \ref{figure3}d. All orange and blue curves in Fig. \ref{figure3} are single-exponential fits (see supplemental material), the fit parameters of which are summarized in Table \ref{table2}. Vertical dashed orange and blue lines indicate the positions $t_\mathrm{ex}$, where the respective pump-probe signals reach their extrema. Note that the fit for the transient band gap in Fig. \ref{figure3}b is sensitive to the value assumed for the equilibrium gap size (we assumed $E_\mathrm{gap}^\mathrm{equ}=2.08\,\mathrm{eV}$) which also affects $\Delta E_\mathrm{charge}^\mathrm{WS\textsubscript{2}}$ in Fig. \ref{figure3}c. We would like to stress, though, that varying the equilibrium gap size by a reasonable amount has only a minor influence on the data points in Fig. \ref{figure3}c. Considering this, the behavior of the transient WS\textsubscript{2} band gap at $k=1.2\,\AA^{-1}$ is very similar for $\hbar\omega=2.0\,\mathrm{eV}$ and $\hbar\omega=3.1\,\mathrm{eV}$. In contrast to this, both the WS\textsubscript{2} and the graphene charging shifts are found to depend on the pump photon energy. We find that both charging shifts reach their extrema $160-340\,\mathrm{fs}$ later for $\hbar\omega=2.0\,\mathrm{eV}$ than for $\hbar\omega=3.1\,\mathrm{eV}$.

\section{Discussion}

The results obtained for $\hbar\omega=2.0\,\mathrm{eV}$ are in good agreement with previous trARPES results \cite{Aeschlimann2020b, Krause2020, Hofmann2023} making their interpretation straight forward. The asymmetric population dynamics of the Dirac cone (Fig. \ref{figure2}a1), the gain above the equilibrium position of the WS\textsubscript{2} VB (Fig. \ref{figure2}b1) and the charging shifts in Figs. \ref{figure3}c and d provide direct evidence of ultrafast charge separation in the WS\textsubscript{2}-graphene heterostructure, where hole transfer from WS\textsubscript{2} to graphene is much faster than electron transfer. The relevant charge transfer states for the WS\textsubscript{2}-graphene heterostructure have been previously identified using density functional theory calculations \cite{Hofmann2023}. For convenience, we sketch the relevant part of the band structure in Fig. \ref{figure5}. Charge transfer states, where the WS\textsubscript{2} and graphene bands hybridize, are shown in red. For $\hbar\omega=2.0\,\mathrm{eV}$, photoexcitation occurs at the K valley of WS\textsubscript{2}. The previously proposed scattering processes resulting in ultrafast charge transfer to the graphene layer are indicated by orange arrows \cite{Krause2020, Hofmann2023}. The transient reduction of the band gap in Fig. \ref{figure3}b has been previously attributed to screening of the Coulomb interaction by photoinduced free carriers \cite{Ugeda2014, Chernikov2015, Ulstrup2016, Pogna2016, Cunningham2017, Lin2022, Hofmann2025}.

The observed differences between excitation at $\hbar\omega=2.0\,\mathrm{eV}$ and $\hbar\omega=3.1\,\mathrm{eV}$, however, deserve further discussion. We start with the observation that, for $\hbar\omega=3.1\,\mathrm{eV}$, the population dynamics in the WS\textsubscript{2} CB in Fig. \ref{figure2}c2 are different at K and close to Q. The $3.1\,\mathrm{eV}$  pump pulse directly populates states close to the Q valley (see Fig. \ref{figure1}a) \cite{Magnozzi2020}. These carriers are then observed to scatter to the K valley within $\sim130\,\mathrm{fs}$  (see Fig. \ref{figure2}c2) in good agreement with previously reported intervalley scattering times for WS\textsubscript{2} and similar TMDCs \cite{Wallauer2016, Bertoni2016, Dong2021a, Wallauer2021, Bange2023}.

Next, we focus on the observation that the gain above the equilibrium position of the WS\textsubscript{2} VB (Fig. \ref{figure2}b) and the charging shifts (Fig. \ref{figure3}c and d) are found to reach their extrema at earlier time delays for $\hbar\omega=3.1\,\mathrm{eV}$ than for $\hbar\omega=2.0\,\mathrm{eV}$. This indicates that hole transfer is faster for $\hbar\omega=3.1\,\mathrm{eV}$ than for $\hbar\omega=2.0\,\mathrm{eV}$. From \cite{Krause2020} it is known that the rates for charge transfer from WS\textsubscript{2} to graphene increase with increasing electronic temperature as it becomes easier for the photoexcited carriers to overcome the energy barriers to the closest charge transfer states. 

The transient electronic temperature for the WS\textsubscript{2} layer is difficult to determine due to the limited signal-to-noise ratio. Therefore, we estimate the peak electronic temperatures for electrons and holes in WS\textsubscript{2} as described in detail in the supplemental material. We obtain $T_\mathrm{max}^{3.1\,\mathrm{eV}}\sim3700\,\mathrm{K}>T_\mathrm{max}^{2.0\,\mathrm{eV}}\sim1000\,\mathrm{K}$ as the excess energy $\frac{1}{2}(\hbar\omega-E_\mathrm{gap})$ of the carriers is considerably larger at $\hbar\omega=3.1\,\mathrm{eV}$. This, indeed, makes it easier for the holes to overcome the energy barrier $\Delta E_\mathrm{K}\sim170\,\mathrm{meV}$ (see Fig. \ref{figure5}). In addition, $T_\mathrm{max}^{3.1\,\mathrm{eV}}$ is high enough for the holes to reach the second charge transfer state at $Q<k<K$ in the VB with an energy barrier of $\Delta E_{QK}\sim410\,\mathrm{meV}$. Hole transfer at $Q<k<K$ is expected to be faster than at $k>K$ because the size of the avoided crossing between WS\textsubscript{2} VB and graphene Dirac cone is bigger (and therefore hybridization is stronger) \cite{Hofmann2023}. The excitation and charge transfer channels for $\hbar\omega=3.1\,\mathrm{eV}$ are indicated by blue arrows in Fig. \ref{figure5}.

Note that the photoexcited carrier densities (an estimate is provided in the Supplementary Material) exceed the Mott threshold \cite{Chernikov2015,Steinhoff2017,Siday2022}, where photoexcited excitons decay into free electron-hole pairs within $\sim100$\,fs \cite{Dendzik2020}. The decay of excitons into free electrons and holes results in a delayed onset of the band gap renormalization in Fig. \ref{figure3}b with respect to the conduction band population in Fig. \ref{figure2}c (see Supplementary Material). However, our data analysis allows us to distinguish between effects caused by screening (transient reduction of the band gap in Fig. \ref{figure3}b) and charge separation (relative shift between the center of the WS$_2$ band gap and the Dirac point in graphene in Figs. \ref{figure3}c and d). Therefore, we believe excitonic effects to be of minor importance for discussing ultrafast charge transfer phenomena in the present work.

Finally, we would like to briefly comment on the possible influence of ultrafast energy transfer between WS\textsubscript{2} and graphene. Following photoexcitation of vdW heterostructures, both charge and energy transfer are possible relaxation channels. While the charging shifts in our data provide direct evidence of ultrafast charge separation and thus ultrafast charge transfer, indications for ultrafast energy transfer are less obvious. Note that our data shows none of the indications of Meitner-Auger energy transfer discussed in \cite{Dong2023}. A possible contribution of F\"orster energy transfer \cite{Tebbe2024} is difficult to evaluate from our data. However, we would like to point out that the presence or absence of additional energy transfer does not affect our conclusions regarding ultrafast charge transfer.

\section{Summary and Outlook}

In summary, we used trARPES of a WS\textsubscript{2}-graphene heterostructure to show that the elevated carrier temperature following C-exciton excitation activates a second, highly efficient charge transfer channel for holes. These results show that charge separation across vdW interfaces can be controlled via the incident pump photon energy, opening up new strategies to optimize the performance of vdW heterostructures for future applications in light harvesting and detection.

\section{Acknowledgments}
This work received funding from the European Union’s Horizon 2020 research and innovation program under Grant Agreement No. 851280-ERC-2019-STG (DANCE) and No. 101130384 (QUONDENSATE), from the German Research Foundation (DFG) via the Collaborative Research Centre 1277 (Project No. 314695032) and the Research Unit 5242 (Project No. 449119662), as well as from the German Federal Ministry of Education and Research (BMBF) (Project No. 05K2022).

\clearpage

\pagebreak

\bibliography{literature_scattering}

\pagebreak
	
	\begin{figure}
		\includegraphics[width = \columnwidth]{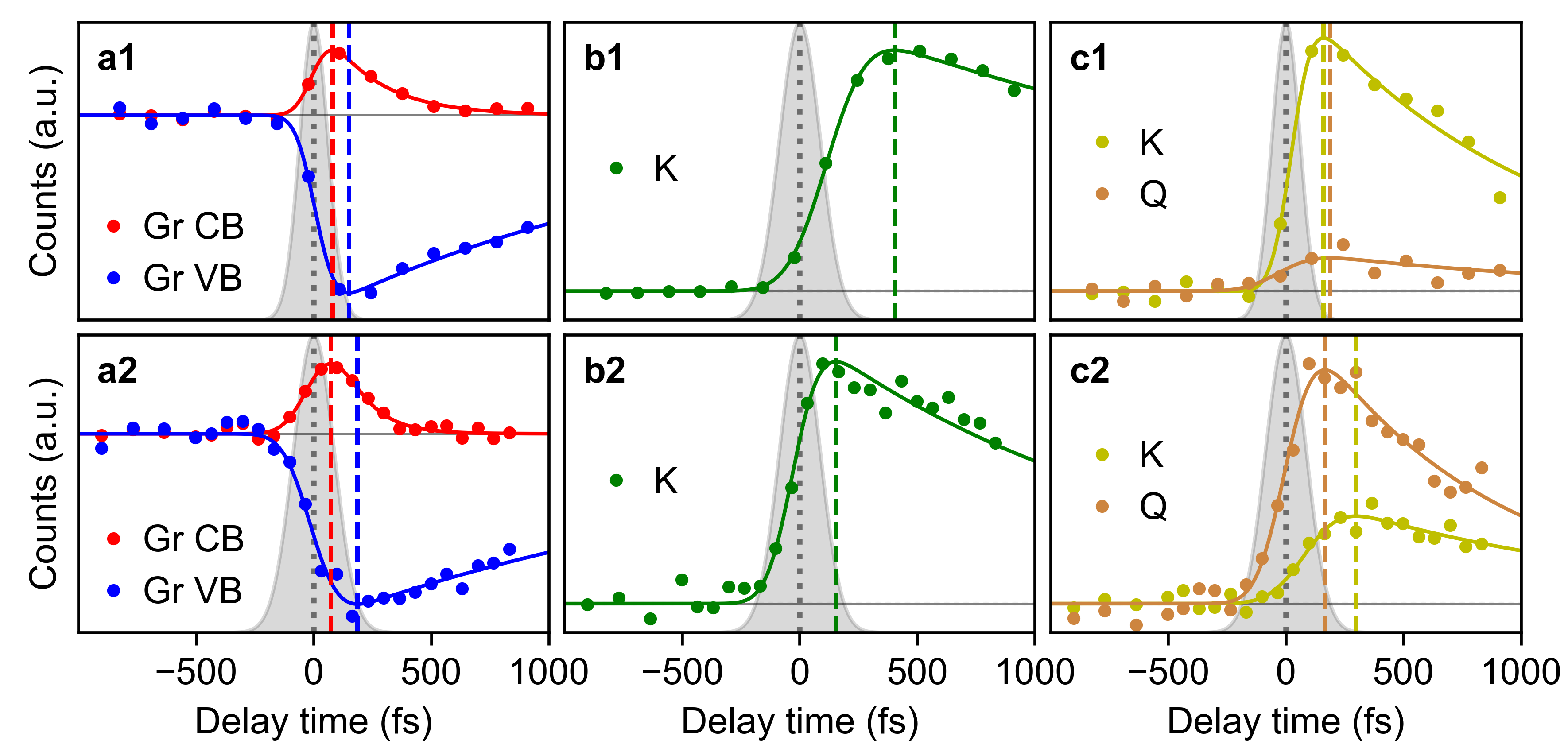}
		\caption{\textbf{Population dynamics} obtained by integrating the counts over the areas marked by colored boxes in Figs. \ref{figure1}b and c. \textbf{a)} Gain (red) and loss (blue) inside the Dirac cone. \textbf{b)} Gain above the equilibrium position of the WS\textsubscript{2} VB at K. \textbf{c)} Population of the WS\textsubscript{2} CB at K (yellow) and close to Q (orange). Rows 1 and 2 present data for excitation at $\hbar\omega=2.0\,\mathrm{eV}$ and $\hbar\omega=3.1\,\mathrm{eV}$, respectively. Continuous lines are single-exponential fits. Dashed colored lines indicate the pump-probe delay where the pump-probe traces reach their respective maximum or minimum. Grey-shaded areas indicate the temporal profile of the pump-probe cross-correlation.}
	\label{figure2}
	\end{figure}
	
	\begin{figure}
		\includegraphics[width = \columnwidth]{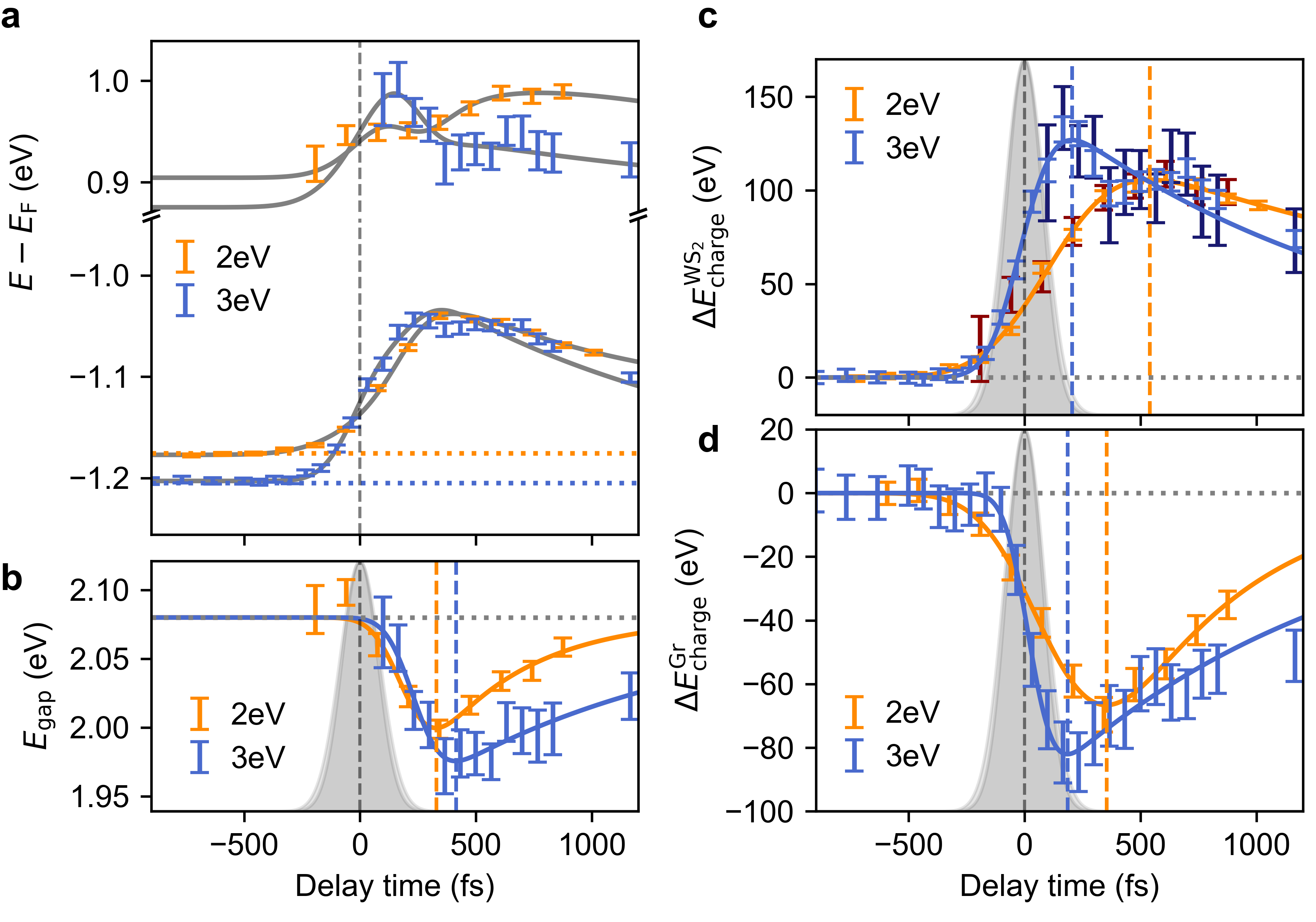}
		\caption{\textbf{Transient band structure.} \textbf{a)} Transient band position of WS\textsubscript{2} CB (top) and VB (bottom).  \textbf{b)} Transient band gap. \textbf{c)} Charging shifts of WS\textsubscript{2} obtained from the VB (blue, orange) and CB (dark blue, red). \textbf{d)} Charging shift of Dirac cone. Data points for excitation at $\hbar\omega=2.0\,\mathrm{eV}$ and $\hbar\omega=3.1\,\mathrm{eV}$ are shown in orange and blue, respectively. Orange and blue curves are single-exponential fits. Grey curves in a) are guides to the eye calculated from the VB fit in a) and the gap fit in b) for the CB, and from the gap fit in b) and the charging shift fit in c) for the VB. Vertical colored dashed lines indicate the pump-probe delay where the pump-probe traces reach their respective maximum or minimum. Grey-shaded areas indicate the temporal profiles of the pump-probe cross-correlations.}
	\label{figure3}
	\end{figure}
	
\begin{figure}
		\includegraphics[width = .5\columnwidth]{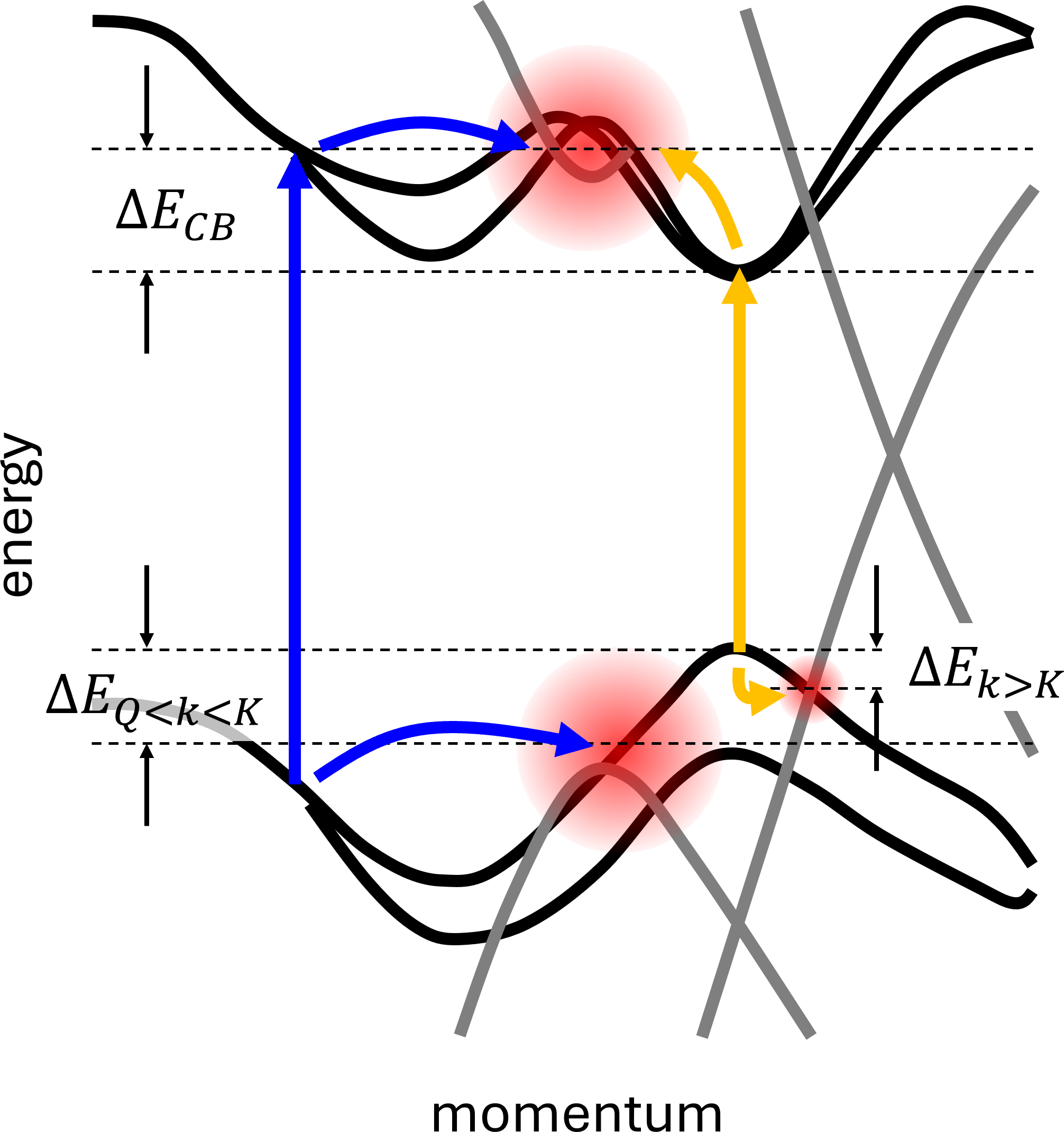}
		\caption{\textbf{Schematic of quantum pathways for ultrafast charge transfer.} The band structure sketch is based on density functional theory calculations from \cite{Hofmann2023}. The color code indicates whether the wave function is localized on the graphene layer (gray curves), the WS\textsubscript{2} layer (black curves) or delocalized over both layers (red dots). Blue (orange) arrows indicate direct optical transitions excited at $\hbar\omega=3.1\,\mathrm{eV}$ ($\hbar\omega=2.0\,\mathrm{eV}$) and subsequent carrier relaxation via different charge transfer states.}
		\label{figure5}
	\end{figure}

\begin{table}[]
	\begin{tblr}{cells=c, rowsep=1pt, rows = {font=\linespread{1.0}\selectfont}}
		\hline
		&Red box &Blue box  &Green box  &Yellow box &Orange box  \\
		\hline
		&{Graphene\\ gain}  &{Graphene\\ loss}  &{Gain above\\ equilibrium position\\ of WS\textsubscript{2} VB}  &{Gain in WS\textsubscript{2}\\ CB at K} &{Gain in WS\textsubscript{2}\\ CB close to Q} \\
		\hline
		$\tau_{2.0\,\mathrm{eV}}$ & $230\pm40$\,fs & $1.7\pm0.2$\,ps & $3.2\pm0.4$\,ps & $1.0\pm0.1$\,ps & $1.3\pm0.8$\,ps \\
		$\tau_{3.1\,\mathrm{eV}}$ & $120\pm30$\,fs & $2.2\pm0.2$\,ps & $1.5\pm0.1$\,ps & $0.8\pm0.1$\,ps & $0.8\pm0.1$\,ps \\
		\hline
		$t_{\mathrm{ex},\,2.0\,\mathrm{eV}}$ & $\sim80$\,fs & $\sim150$\,fs & $\sim400$\,fs & $\sim160$\,fs & $\sim190$\,fs \\
		$t_{\mathrm{ex},\,3.1\,\mathrm{eV}}$ & $\sim70$\,fs & $\sim190$\,fs & $\sim160$\,fs & $\sim300$\,fs & $\sim170$\,fs \\
		\hline
	\end{tblr}
	\caption{Fit parameters for single-exponential fits of pump-probe traces shown in Fig. \ref{figure2}.}
	\label{table1}
\end{table}

\begin{table}[]
	\begin{tblr}{cells=c, rowsep=3pt}
		\hline
		&WS\textsubscript{2} VB position &$E_\mathrm{gap}$ &$\Delta E_\mathrm{charge}^\mathrm{WS\textsubscript{2}}$ &$\Delta E_\mathrm{charge}^\mathrm{Gr}$  \\
		\hline
		$\tau_{2.0\,\mathrm{eV}}$ & $1.5\pm0.2$\,ps & $0.4\pm0.2$\,ps & $2.7\pm0.9$\,ps & $0.6\pm0.1$\,ps \\
		$\mathrm{Amplitude}_{2.0\,\mathrm{eV}}$ & $137\pm6$\,meV & $80\pm30$\,meV & $110\pm10$\,meV & $70\pm10$\,meV \\
		\hline
		$\tau_{3.1\,\mathrm{eV}}$ & $1.3\pm0.1$\,ps & $1.2\pm0.2$\,ps & $1.5\pm0.2$\,ps & $1.3\pm0.1$\,ps \\
		$\mathrm{Amplitude}_{3.1\,\mathrm{eV}}$ & $169\pm8$\,meV & $104\pm8$\,meV & $127\pm6$\,meV & $82\pm3$\,meV \\
		\hline
		$t_{\mathrm{ex},\,2.0\,\mathrm{eV}}$ & $\sim350$\,fs & $\sim330$\,fs & $\sim540$\,fs & $\sim350$\,fs \\
		$t_{\mathrm{ex},\,3.1\,\mathrm{eV}}$ & $\sim400$\,fs & $\sim410$\,fs & $\sim200$\,fs & $\sim190$\,fs \\
		\hline
	\end{tblr}
	\caption{Fit parameters for single-exponential fits of pump-probe traces shown in Fig. \ref{figure3}.}
	\label{table2}
\end{table}

\clearpage
\pagebreak

\par\ \ \vspace{0pt}\par\ \ 
	{\centering\bfseries\LARGE SUPPLEMENTAL MATERIAL\\}

	\section{Exponential decay fits}
	
	To extract rise and decay times of a given pump-probe signal, we use the following analytic fitting function that stems from convolving a step function with exponential decay for the underlying dynamics with a Gaussian accounting for the finite temporal resolution:
	
	\begin{equation}
		f(t) = \frac{a}{2} \left( 1 + \erf\left( \frac{(t-t_0)\tau - \frac{\mathrm{FWHM}^2}{8\ln2}}{\sqrt{2}\tau\frac{\mathrm{FWHM}}{2\sqrt{2\ln2}}} \right) \right)
		\exp\left( \frac{\frac{\mathrm{FWHM}^2}{8\ln 2}-2(t-t_0)\tau}{2\tau^2} \right).
	\end{equation}

	Here, $a$ is the amplitude, $\erf$ is the error function, $\mathrm{FWHM}$ is the full width at half maximum of the Gaussian, $t_0$ is the time delay where pump and probe overlap, and $\tau$ is the decay time. This function was used to fit the time traces in Figs. 2 and 3 in the main manuscript.

	\section{Extracting transient peak positions}
	
\begin{figure}[]
    \centering
    \includegraphics[width=\linewidth]{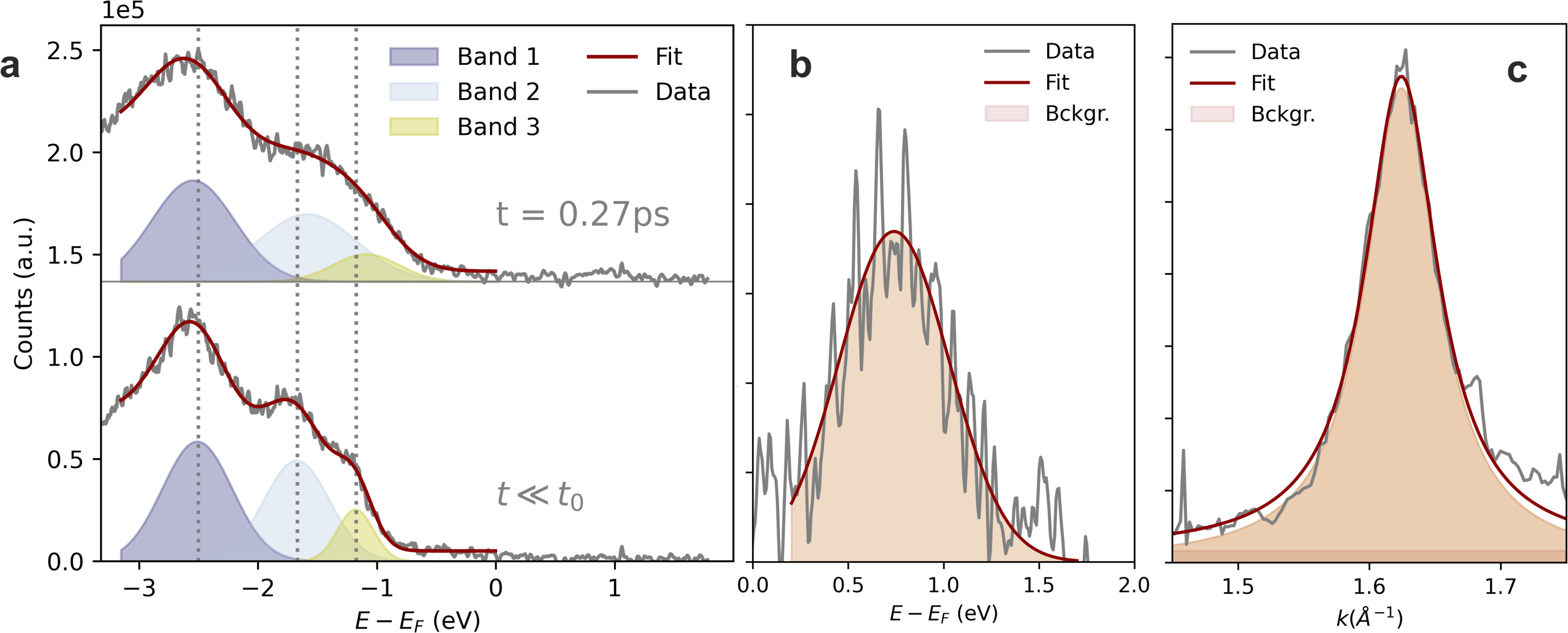}
    \caption{\textbf{Extracting transient peak positions. a)} WS\textsubscript{2} EDCs for negative pump-probe delay and at $t=\SI{0.2}{\pico\second}$ fitted with a constant offset and three Gaussian peaks. The dotted vertical lines mark the peak positions for negative pump-probe delay. \textbf{b)} Pump-induced changes of the EDC through the WS\textsubscript{2} conduction band together with a Gaussian fit. \textbf{c)} MDC through the graphene Dirac cone at negative pump-probe delay fitted with the sum of a constant offset and a Lorentzian peak.}
		\label{figureS1}
\end{figure}

	For extracting the transient position of the WS\textsubscript{2} valence band, the data in Fig.\,\ref{figureS1}a (main manuscript) were integrated over the momentum range $\Delta k=\pm\SI{0.04}{\per\angstrom}$ for $2.0\,\mathrm{eV}$ and $\Delta k=\pm\SI{0.06}{\per\angstrom}$ for $3.1\,\mathrm{eV}$ around $k=\SI{1.2}{\per\angstrom}$. The latter was chosen to ensure that the graphene valence band does not influence the data. The resulting energy distribution curves (EDCs) were then fitted with a constant background and three Gaussian peaks for the three expected bands. Exemplary fits are presented in Fig.\,\ref{figureS1}a. The following constraints were applied:
	\begin{itemize}
		\item The constant background was fixed to the value found for negative pump-probe delay.
		\item The energy difference between the two spin-split WS\textsubscript{2} valence bands was fixed to the value found for negative pump-probe delay.
		\item The spectral weight of bands 1 and 2 was fixed to the value found for negative pump-probe delay.
	\end{itemize}
	
	To obtain the transient position of the WS\textsubscript{2} conduction band, the data in Figs.\,1b and c (main manuscript) were integrated over the momentum range $\Delta k=\pm\SI{0.1}{\per\angstrom}$ around $k=\SI{1.3}{\per\angstrom}$. The resulting EDCs were then fitted with a background-free Gaussian (see Fig.\,\ref{figureS1}b).\\
	
	The shift of the graphene Dirac cone was determined by integrating the data in Fig. 1a (main manuscript) over the energy range $\Delta E = \pm\SI{25}{\milli\electronvolt}$ around the five energy positions $(E-E_F)= \SI{-1.0}{\electronvolt},\SI{-0.8}{\electronvolt},\SI{-0.6}{\electronvolt},\SI{-0.4}{\electronvolt},\SI{-0.2}{\electronvolt}$. To improve the signal-to-noise ratio, we grouped the resulting momentum distribution curve (MDC) data into bins of three consecutive time delays. The final MDCs were fitted by the sum of a constant background and a Lorentzian peak (see Fig.\,\ref{figureS1}c) for every energy position. The five momentum shifts obtained in this way were averaged and then converted into an energy shift by multiplying with the slope of the $\pi$-band of \SI{7}{\electronvolt\angstrom}.
	
	\section{Estimating transient carrier temperatures in WS\textsubscript{2}}
	
	According to \cite{Erben2022}, the absorption in the relevant fluence regime is highly non-linear. The excited carrier density for resonant excitation of the C-exciton with a fluence of $0.4\,\mathrm{mJ/cm}^2$ can be directly obtained from \cite{Erben2022} as $n_\mathrm{ex}\sim1\times 10^{14}\,\mathrm{cm}^{-2}$. \cite{Erben2022} also showed that absorption at the A-exciton resonance drops to zero when similar carrier densities are reached. Therefore, we assume that both $2.0\,\mathrm{eV}$ and $3.1\,\mathrm{eV}$ excitation generate carrier densities on the order of $n_\mathrm{ex}\sim1\times 10^{14}\,\mathrm{cm}^{-2}$ even though the incident fluences differ considerably.\\
	
	To estimate the peak carrier temperature, we first assume that the excess energy for electrons and holes $E_\mathrm{excess}\sim\frac{1}{2}\left(\hbar\omega_\mathrm{pump}-E_\mathrm{gap}\right)$ can be directly converted into a carrier temperature $k_BT$. The excess energy for $3.1\,\mathrm{eV}$ excitation is $500\,\mathrm{meV}$ yielding $T_{3.1\,\mathrm{eV}} \sim 5800\,\mathrm{K}$. The A-exciton resonance, however, lies below the single-particle gap. In our case, the photoexcited carrier densities exceed the Mott threshold, such that the excitons rapidly decay into free electron-hole pairs. These free electron-hole pairs screen the Coulomb interaction which results in a transient reduction of the single particle band gap by $\sim80\,\mathrm{meV}$ (see Fig. 3b of main manuscript). Therefore, it seems justified to assume that, for $2.0\,\mathrm{eV}$ excitation, electrons and holes gain an excess energy of $\sim40\,\mathrm{meV}$ yielding a transient carrier temperature of  $T_{2.0\,\mathrm{eV}} \sim 500\,\mathrm{K}$.\\
	
	An improved estimate can be obtained by calculating the carrier temperature in the framework of the quasi-free electron model. Here, the change of the internal energy due to heating of a given carrier density $n_\mathrm{ex}$ from room temperature to an arbitrary temperature $T$ is given by 
	
	$$\Delta U = \int\displaylimits^\infty_{E_\mathrm{CB}=0} E\cdot\mathrm{DOS}(E)\cdot f_\mathrm{FD}(E, \mu, T)\,\mathrm{d}E - \int\displaylimits^\infty_{E_\mathrm{CB}=0} E\cdot\mathrm{DOS}(E)\cdot f_\mathrm{FD}(E, \mu, T=300K)\,\mathrm{d}E,$$
	
	where the 2D $\mathrm{DOS}(E) = \frac{m^*g_sg_v}{2\pi\hbar^2}$. $g_s=2$ and $g_v=2$ are the spin and valley degeneracies, respectively. From the carrier concentration $n_\mathrm{ex}$ we obtain the chemical potential $\mu=\frac{2\pi\hbar^2}{m^*g_sg_v}n_\mathrm{ex}$ that, in Sommerfeld approximation, is independent of temperature.	The part of the absorbed fluence that is available for heating the carriers is given by the product of the photoexcited carrier density $n_\mathrm{ex}$ and the excess energy $E_\mathrm{excess}$. From $n_\mathrm{ex}E_\mathrm{excess}=\Delta U$ we obtain $T_{2.0\,\mathrm{eV}} \sim 1000\,\mathrm{K}$ and $T_{3.1\,\mathrm{eV}} \sim 3700\,\mathrm{K}$ for an effective mass of $m^*=0.5m_e$.

\section{Decay of excitons into free electrons and holes}

\begin{figure}[h]
	\centering
	\includegraphics[width=\linewidth]{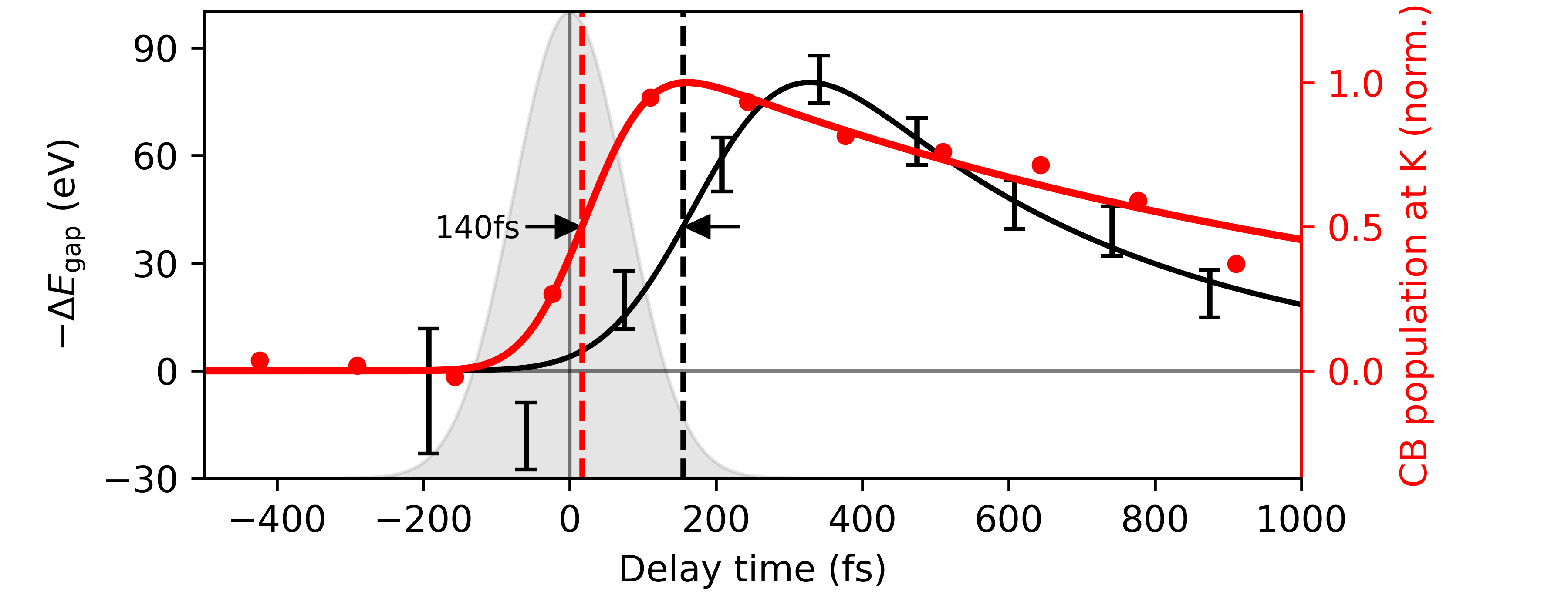}
	\caption{\textbf{Excitonic effects.} Direct comparison between transient band gap reduction $-\Delta E_\mathrm{gap}$ (black) and conduction band population (red) for resonant excitaton of the A-exciton. The offset between the two rising edges of $\sim140$\,fs is caused by the decay of excitons into free electrons and holes.}
	\label{figureS2}
\end{figure}

The estimated carrier densities on the order of $n_\mathrm{ex}\sim1\times 10^{14}\,\mathrm{cm}^{-2}$ exceed the Mott threshold and are expected to decay into free electrons and holes within $\sim100$\,fs \cite{Dendzik2020}. Free charge carriers efficiently screen the Coulomb interaction and are responsible for the transient reduction of the band gap reported in Fig. 3b (main manuscript). The decay of excitons into free electrons and holes shows up as a delayed onset of the band gap renormalization with respect to the transient population of the conduction band in Fig. \ref{figureS2}.

\bigskip

\vfill
	
%

	\end{document}